\documentclass[twocolumn]{openjournal}

\usepackage{newtxtext,newtxmath}
\usepackage[T1]{fontenc}

\makeatletter
\def\fnum@table{{\@eapj@cap@font Table~\thetable.}}
\long\def\@makecaption#1#2{%
 \noindent\begin{minipage}{0.9999\linewidth}
   \if\csname ftype@\@captype\endcsname 2 
   \vskip 2ex\noindent \centering\footnotesize{#1}~#2\par\medskip
   \else
   \vspace*{\abovecaptionskip}\noindent\footnotesize #1 #2\par\vskip \belowcaptionskip
   \fi
 \end{minipage}\par
}
\makeatother

\usepackage{graphicx} 
\usepackage{amsmath}
\usepackage{listings}
\usepackage[dvipsnames]{xcolor}
\usepackage{hyperref}
\usepackage{orcidlink}
\usepackage{soul}
\usepackage{lineno}
\usepackage{needspace}

    [2021/06/11 v1.0.4 Linked ORCiD logo macro package]

\begin{document}
\title{Galactic Constellations in DESI DR1 and the Scales of Cosmological Homogeneity}
\author{Claire Lamman\orcidlink{0000-0002-6731-9329},$^{1,2,3}$}
\affiliation{$^1$ Center for Cosmology and AstroParticle Physics (CCAPP), Ohio State University, Columbus, OH 43210}
\affiliation{$^2$ Department of Astronomy, The Ohio State University, Columbus, OH 43210, USA}
\affiliation{$^3$ Department of Physics, The Ohio State University, Columbus, OH 43210, USA}

\begin{abstract}
We present galactic constellations: charming shapes in large cosmological surveys. By exploring a dense subset of DESI's first data release, we discover distinctive constellations including \textit{Pisces Grandis}, \textit{The DESI Stick Woman}, and \textit{W}. We additionally develop a public website for anyone to explore DESI data, find their own constellations, and share their creations: \href{https://cmlamman.github.io/galactic-constellations/index.html}{cmlamman.github.io/galactic-constellations}. Early users of the site discovered 93 constellations. We analyze the size of these constellations as an unconventional probe of homogeneity, finding consistency with the cosmological principle and $\Lambda$CDM.

\end{abstract}

\section{Introduction}
Humans have recorded stellar constellations for at least the past 5,000 years \citep{rogersOriginsAncientConstellations1998}. Within the past 50, we have developed the ability to map out the positions of galaxies and unleashed a new frontier of finding patterns in the sky. 

\begin{figure}[t]
\centering
\includegraphics[width=.49\textwidth]{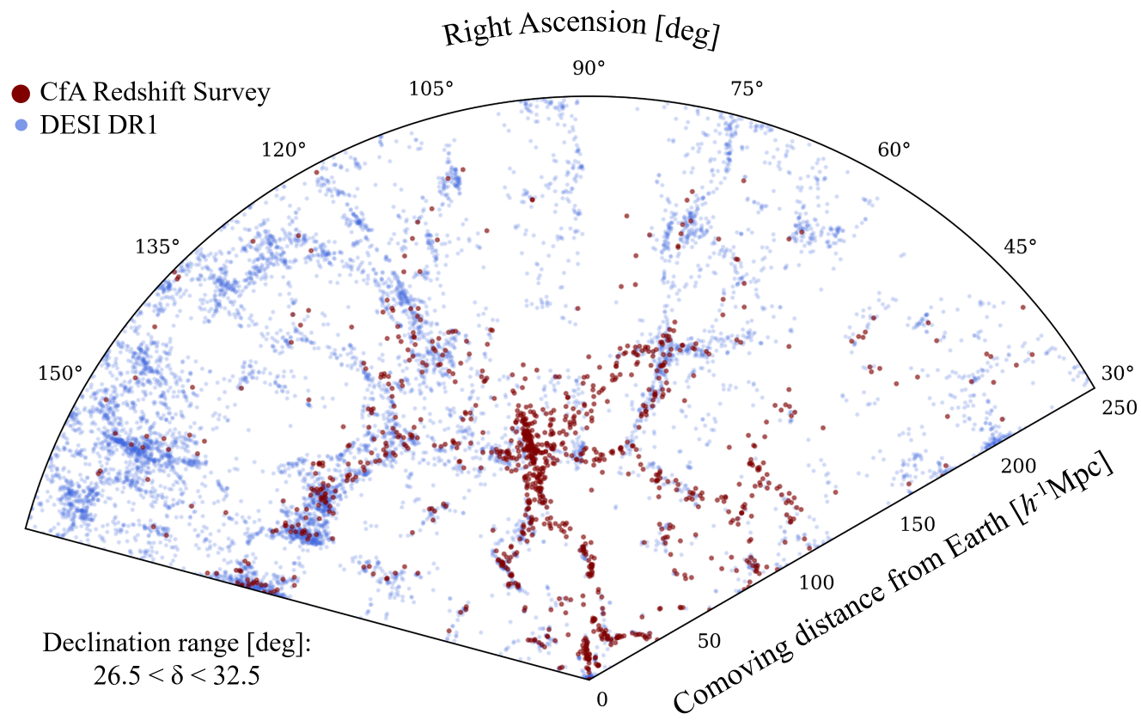}
\caption[]{\textit{The CfA Stick Man}, the first galactic constellation. It was discovered in the CfA Redshift Survey, shown in red. Smaller blue points in the background from DESI demonstrate that he is not alone.}
\label{fig:stickman}
\end{figure}

The earliest distinctive galactic constellation was found in observations from 1985, as part of the historic CfA Redshift Survey \citep{huchraCfARedshiftSurvey1990}. Spectroscopic measurements of galaxies within the Coma Cluster revealed the form of a flailing figure, now known as \textit{The CfA Stick Man}. He is shown in Figure \ref{fig:stickman}, outlined in both the originally observed galaxies and new measurements from the Dark Energy Spectroscopic Instrument (DESI). 

\textit{The CfA Stick Man} represents the first conclusive evidence that galaxies are not randomly scattered about, but trace the large-scale structure (LSS) of the universe. We are surrounded by a massive web of dark matter that is illuminated by sheets, voids, and clusters of galaxies. Measuring these large-scale patterns informs us about the forces which form them: gravity and dark energy. Mapping LSS has been a priority within cosmology, and DESI is five years into an eight-year program to map out the positions of 63 million galaxies \citep{Snowmass2013.Levi}. Early results put dark energy to its strictest tests yet and indicate a not-so-constant cosmological constant \citep{DESI2024.VII.KP7B, abdulkarimDESIDR2Results2025}. 

As we map the cosmic web in increasing detail, cosmologists develop increasingly complicated statistics to capture all its information. Here we explore the highest-order statistic yet: galactic constellations. Patterns that vaguely resemble familiar objects are difficult to study objectively and challenging to tie to cosmological parameters. However, they are fun. Identifying relatable shapes in galaxy maps can help people develop an intuition for characteristic patterns and the scales at which they are present. It is also an engaging way for general audiences to interact with data and become introduced to the ideas of galaxy surveys and LSS. 

In this work we present the discovery of several new galactic constellations in DESI's first public data release. We also develop a website that allows people to discover and share their own constellations. This makes DESI's map more functionally accessible, giving anyone the opportunity to find a new pattern yet unseen by human eyes. Finally, we explore using the sizes of discovered constellations to probe the scales of  cosmological homogeneity.


\section{DESI Data}\label{sec:data} 
DESI is an instrument on the 4-meter Mayall telescope on Kitt Peak \citep{DESI2016b.Instr, desicollaborationOverviewInstrumentationDark2022}. It can gather thousands of spectroscopic redshifts in minutes with 5000 autonomous robots that position fiber optic cables onto galaxies and feed their light to a set of spectrographs  \citep{millerOpticalCorrectorDark2024, poppettOverviewFiberSystem2024, guySpectroscopicDataProcessing2023}. Despite COVID-19, a wildfire, and the ladybug swarms, DESI has been reliable and efficient \citep{schlaflySurveyOperationsDark2023}. After recently completing its originally-planned survey, DESI is now continuing through 2028 and will gather 63 million total extragalactic redshifts, covering 17,000 deg$^2$ \citep{bironDESICompletes2026}. DESI's First Data Release (DR1) is public and contains 14 million extragalactic spectra \citep{DESI2024.I.DR1}. 

\begin{figure*}
\centering
\includegraphics[width=.98\textwidth]{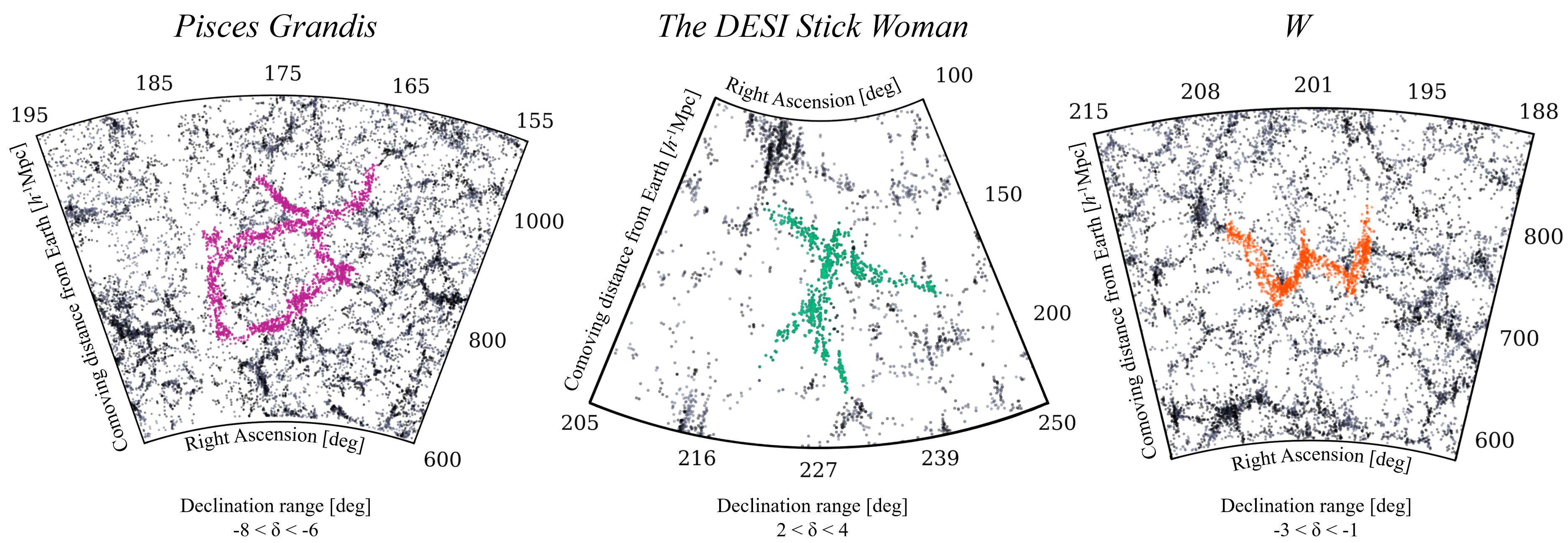}
\caption[]{A set of especially distinctive constellations found in DESI's map of galaxies, highlighted in color on each panel. Left: \textit{Pisces Grandis} swimming towards Earth. At 270 $h^{-1}$Mpc long, this is the largest known fish in the Universe. Middle: \textit{The DESI Stick Woman}, a potential friend for \textit{The CfA Stick Man}. Right: \textit{W}. Astronomers may note a similarity to the stellar constellation \textit{Cassiopeia}. All were found within DESI's DR1 in 2-degree slices of declination.}
\label{fig:constellations}
\end{figure*}

We focus our constellation search on physical patterns; ones most likely to reflect the true underlying distribution of dark matter that will not dramatically change as more galaxies are observed. For this, we cut the DR1 data to a patch of sky with the highest survey completeness, lying between $125 <\alpha < 250$ degrees in right ascension and $-7 < \delta < 7$ degrees in declination. Therefore, most of our galactic constellations lie in the stellar constellation \textit{Virgo}. We also limit to the most dense region of the survey: redshifts below $z<0.8$. This includes a portion of DESI's Luminous Red Galaxy (LRG) sample \citep{zhouTargetSelectionValidation2023}, but is mostly made of the Bright Galaxy Survey (BGS) sample \citep{hahnDESIBrightGalaxy2023}. This constitutes the most detailed map of the nearby universe available, and LSS is clearly visible in slices from this complete section.

This data also contains visually recognizable evidence of redshift-space distortions (RSD). Although these patterns are a departure from the galaxies' true positions, RSD is a scientific and historical feature of these maps. Without RSD, the stick man would not be quite as tall. Future work could include reconstructed maps, but accuracy would be poor for individual constellations.

\section{Finding Constellations}
\subsection{Initial Search}
To visually identify distinctive features of the cosmic web, we split the 3-dimensional data in slices of varying size. Since the geometry of the sky selection is wider in right ascension, our slices were made in declination. We found that a slice thickness of 2 degrees is best for distinguishing large physical structures without overcrowding the field. Although we created 2-dimensional images, the brightness of plotted points was scaled by declination to distinguish foreground and background structure. The galaxies were also plotted with semi-transparent points to visually emphasize over-dense regions. All constellations were found in comoving space. Galaxy comoving positions were determined from their measured redshifts and cosmological parameters from Planck 2018 \citep{collaborationPlanck2018Results2020}.

The initial search was done by a biological neural network (\textbf{C}\textsc{omplex} \textbf{L}\textsc{earning} \textbf{A}\textsc{nd} \textbf{I}\textsc{ntrospective} \textbf{R}\textsc{eality} \textbf{E}\textsc{ngine}), equipped with 29 years of real-world training data. This follows recently proposed methods for deeper learning in astronomy \citep{scottDeeperLearningAstronomy2024}.
A set of the first constellations discovered are shown in Figure \ref{fig:constellations}. They include \textit{Pisces Grandis}, \textit{The DESI Stick Woman}, and \textit{W}. We believe \textit{Pisces Grandis} to be the largest known fish in the universe, at 270 $h^{-1}$Mpc long, or about 900 million light years. And he is only getting bigger. An observer inside \textit{Pisces Grandis} will see him expanding from his tip to his tail at a rate of 10,000 miles per second.

\vspace{-2mm}
\subsection{Interactive Website}
To find more constellations and enable anyone to discover them, we developed a simple interactive website, \href{https://cmlamman.github.io/galactic-constellations/index.html}{cmlamman.github.io/galactic-constellations}. This contains two main features: a constellation search tool and a sharing / voting system.

The search tool displays a random slice from the DESI DR1 region described in Section \ref{sec:data}, each with a width of 2 degrees in declination. As above, point colors and transparencies are scaled to highlight depth and overdensities. They are cut to 509 pre-generated square images with side lengths that range from 100-1000 $h^{-1}$Mpc. Images with large artificial gaps from the survey boundary were removed. A new image is displayed upon each page refresh with random orientation. The user can rotate the image or move to a new one. After drawing their constellation on the galaxy wedge, they can download or share their discovery. Submitted constellations are saved to the site, along with their name, the name of the founding astronomer, and details about the location and size of the galaxy slice.

\begin{figure*}
\centering
\includegraphics[width=.9\textwidth]{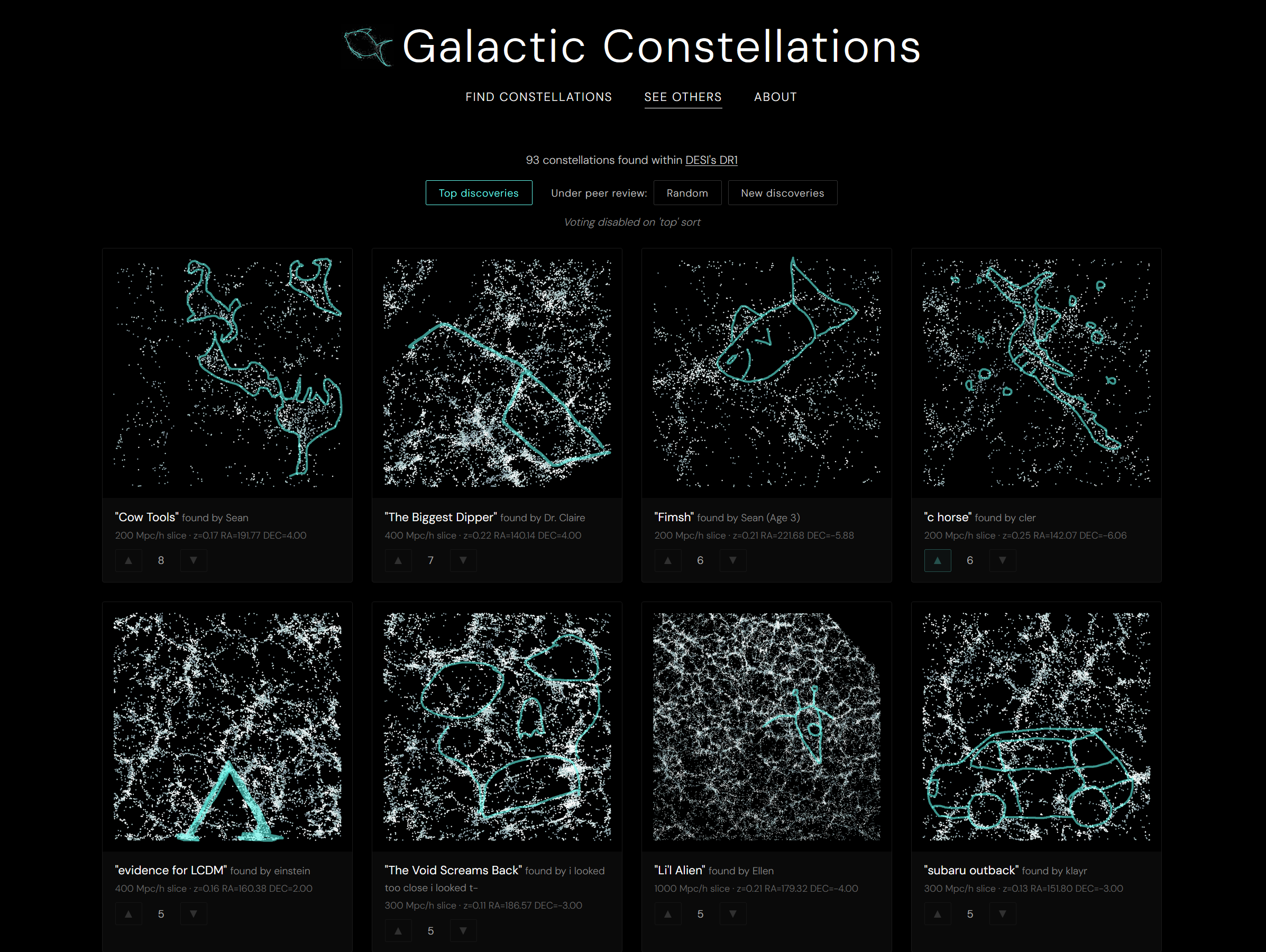}
\caption[]{A screenshot of the gallery page from our \href{https://cmlamman.github.io/galactic-constellations/index.html}{public website}, where users can find their own galactic constellations. After drawing and submitting constellations under "Find Constellations", the artwork is displayed on the site for others to see and vote on. These represent the top submissions from an early pool of users.}
\label{fig:website}
\end{figure*}

The submitted constellations are displayed on a gallery page. They can be sorted by ``Top discoveries",  ``Random", or ``New discoveries". Users can vote on their favorites under the last two options, which are considered to be under peer review. The mechanisms, clarity, and security of the site were assessed by a set of 20 beta testers, of which half were astronomers. A screenshot of the top discoveries from these early users is shown in Figure \ref{fig:website}. The most popular constellation is \textit{Cow Tools}, a reference to a non sequitur comic and internet meme \citep{CowTools1982}. \textit{Cow Tools} demonstrates the profound and immutable nature of constellations beyond geometric shapes: they are cultural artifacts that encode the shared mythology of the civilization that names them.

\subsection{A note on Artificial Intelligence}\label{sec:AI}
We explored the use of AI to generate constellations, which failed spectacularly. Human brains are uniquely adapted to quickly identify patterns from complex data \citep{mattsonSuperiorPatternProcessing2014}, and still significantly outperform artificial intelligence in learning efficiency \citep{holzingerHumanlevelConceptLearning2023}, visual processing \citep{wichmannAreDeepNeural2023}, and recognizing objects in unusual poses \citep{ollikkaComparisonHumansAI2024}. We shared a set of the website's galaxy images with Claude's Opus 4.6 model. Even with detailed prompting and examples, it struggled to identify recognizable shapes. Although results would improve with a model trained for this specific goal, creatively synthesizing data remains a uniquely human capability.

\section{Cosmological Implications}

The cosmological principle (CP) is a core feature of the standard model and states that the universe is homogeneous and isotropic on large scales. This justifies the use of the FLRW metric to describe space-time and its expansion, which underpins all of standard cosmology \citep{weinbergGravitationCosmologyPrinciples1972}. Recent tensions within the $\Lambda$CDM model motivate a reexamination of the CP \citep{kumaraluriObservableUniverseConsistent2023}, and the scales at which the universe conforms to the CP is sensitive to cosmology, gravity, and astrophysics \citep{ntelisExploringCosmicHomogeneity2017, migkasCosmologicalImplicationsAnisotropy2021, avilaHomogeneityScaleGrowth2022}. However, defining the scales of this transition is nontrivial \citep{wuLargescaleSmoothnessUniverse1999}. The works cited here report theoretical and observational values, with a variety of definitions that place the transition at various points between 60-500 $h^{-1}$Mpc.

DESI's measurements mark a significant advancement in the resolution of cosmic web maps and the complexity of statistics needed to quantify all its information. As described in Section \ref{sec:AI}, humans are uniquely adapted to efficiently pick out complex patterns. Although the identification of individual constellations is subjective, we can examine the scales at which these patterns are identified.

\begin{figure}[h]
\centering
\includegraphics[width=.47\textwidth]{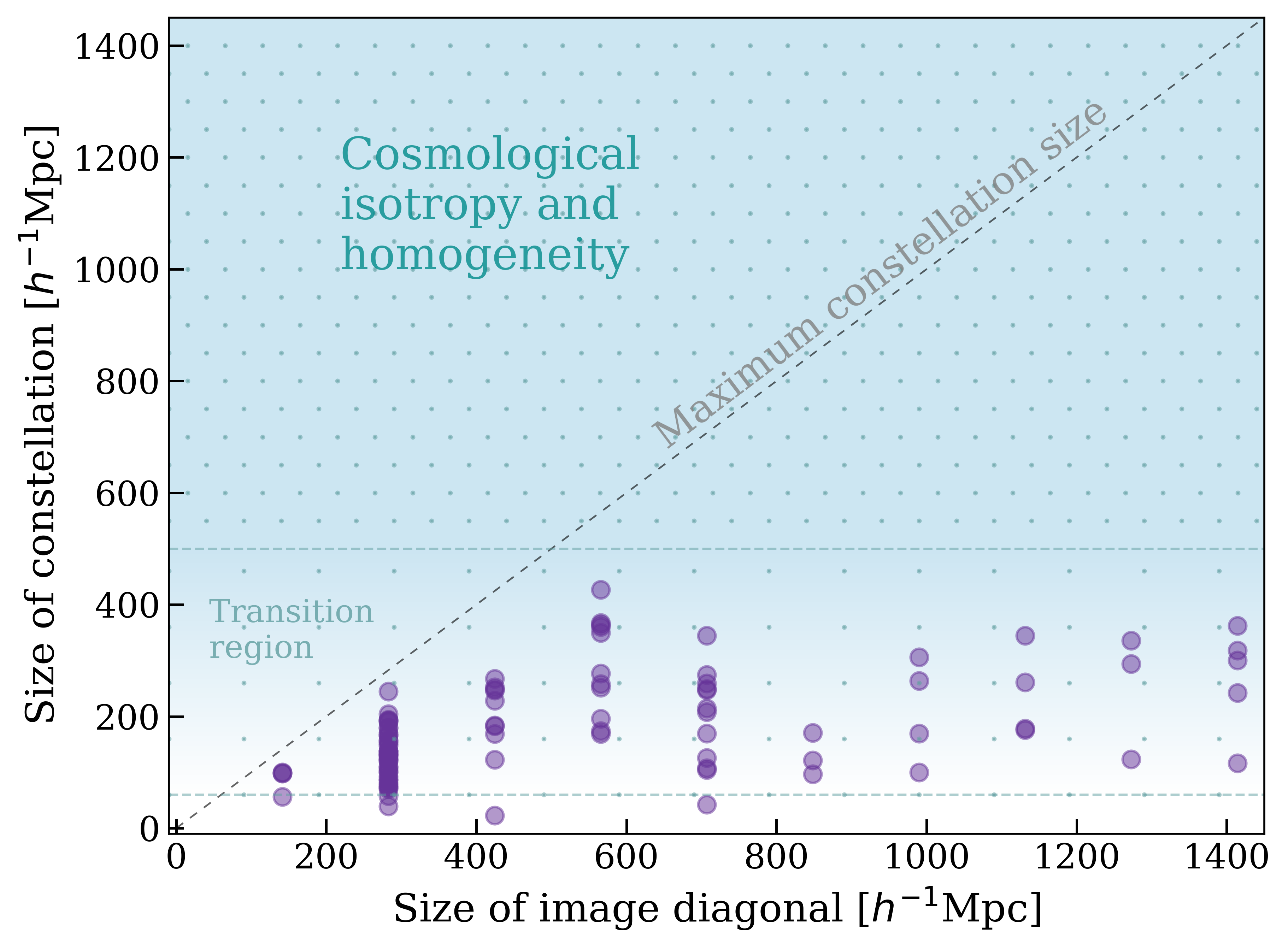}
\caption[]{The largest distance spanned by each constellation drawing, compared to the size of the slice it was identified in. The dotted blue background corresponds to the scales we expect the cosmic web to become isotropic and homogeneous. This ranges from 60 - 500 $h^{-1}$Mpc, depending on definition and method. Cohesive shapes were all identified at scales below the maximum estimates of this transition, consistent with the cosmological principle and $\Lambda$CDM.}
\label{fig:sizes}
\end{figure}

We measure the sizes of the 93 constellations submitted to the website as the maximum distance between points in each drawing, as determined from pixel color. This is compared to the total image size and scaled by the physical span of the image. Although users are shown images of the cosmic web that range in size from 100-1000 $h^{-1}$Mpc, the sizes of constellations they identify are consistently less than 400 $h^{-1}$Mpc (Figure \ref{fig:sizes}). In an anisotropic or inhomogeneous universe, large-scale coherent structures could persist to scales where humans would still identify them as distinctive shapes, so the absence of recognizable patterns above this scale is qualitatively consistent with expectations from the CP. This is an unexpected result of the website constellations; its users were unaware of the cosmological implications of their work.

Robust cosmological inference would require a more uniform distribution of the image sizes available, a larger sample size, and mock data. Future studies could involve identifying constellations in a mix of real and simulated data. A suitable sample size could reveal statistically significant differences between different cosmologies in the features and scales most commonly picked out by humans.

Additionally, we note that no $\Lambda$CDM cosmological simulations have reproduced these exact constellations. For example, out of the 25 mock DESI datasets produced by the A\textsc{bacus} simulation suite \citep{garrisonAbacusCosmosSuite2018, maksimovaAbacusSummitMassiveSet2021}, no large fish has been found. Under a frequentist framework, this implies that \textit{Pisces Grandis} is in tension with $\Lambda$CDM at 96\% confidence. This leads to a striking, inevitable conclusion: frequentist statistics should not be applied to individual structures in cosmological analysis.

\section{Conclusion}
We identify optimal strategies for visualizing a dense subset of DESI's first data release in order to identify distinctive shapes tied to the underlying dark matter distribution. We present the first three galactic constellations found in DESI, \textit{Pisces Grandis}, \textit{The DESI Stick Woman}, and \textit{W}. By developing and sharing a \href{https://cmlamman.github.io/galactic-constellations/index.html}{public website}, we made constellation-finding accessible to anyone and gathered 93 additional constellations. The sizes of discovered constellations are all on scales consistent with expectations from the cosmological principle and $\Lambda$CDM, regardless of the box size they were identified in.

Although not visible to the human eye, galactic constellations have some advantages over stellar ones. The movements of galaxies happen on much longer time scales than stellar orbits within the Milky Way, so galactic constellations will last millions of years longer. Additionally, traditional constellations are only visible to inhabitants of our solar system; galactic ones are universal for any intelligent life within a few billion light years. 

Finally, humans have been lovingly cataloging the same night sky for generations but seem to be running out of ideas \citep{lundAstronomersGettingLess2025}. Galactic constellations are an opportunity to form new figures and stories. The galaxies featured here are only the start. Our galaxy sample represents only 10\% of the sky area that DESI will observe and less than 0.01\% of its total volume. Also, due to the geometry of DR1, we only took slices in declination. Slices in right ascension or the plane of the sky will triple the perspectives to look from, and perhaps we will even find structures in full 3-dimensional space. The website will be updated as additional data becomes available. For now, we encourage the reader to explore the newest data from DESI, search for their own constellations, and continue one of humanity's oldest traditions.

\section*{Acknowledgements}
Thank you to all of the early users of the galactic constellation site. We are especially grateful to Sean Moss and Mathew Salo for extensive testing. Thank you also to esteemed peer reviewers: Ellen Price, Michael Foley, and Theron Carmichael. 

CL is supported by a NSF Postdoctoral Fellowship under award 2502789.

\section*{Data Availability}
DESI DR1 is available at
\href{https://data.desi.lbl.gov/doc/releases/dr1/}{data.desi.lbl.gov/doc/releases/dr1}.

\vspace{.2in}
\bibliographystyle{mnras}
\bibliography{references} 

\end{document}